\documentclass[]{aa}
\usepackage{psfig}

\include{psfig}

\begin{document}


\thesaurus{06(02.04.2; 08.08.1; 08.04.1; 08.01.1; 10.07.2)}


\title{Atomic diffusion in metal poor stars.} 
\subtitle{The influence on the Main Sequence fitting distance scale,
subdwarfs ages and the value of $\Delta Y/ \Delta Z$
}

\author{M. Salaris\inst{1,2}, M.A.T. Groenewegen\inst{2} 
\and A. Weiss\inst{2}
}
\offprints{M. Salaris~(ms@staru1.livjm.ac.uk)}
\institute {
Astrophysics Research Institute, Liverpool John Moores
University, Twelve Quays House, Egerton Wharf, Birkenhead CH41 1LD, UK
\and
Max-Planck-Institut f\"ur Astrophysik, 
Karl-Schwarzschild-Stra{\ss}e 1, D-85740 Garching, Germany
}

\authorrunning{M. Salaris et al.}
\titlerunning{Atomic diffusion in metal poor stars}

\maketitle

\begin{abstract}

The effect of atomic diffusion on the Main Sequence (MS) of metal-poor
low mass stars is investigated.  Since diffusion alters the stellar
surface chemical abundances with respect to their initial values, one
must ensure -- by calibrating the initial chemical composition of the
theoretical models -- that the
surface abundances of the models match the observed ones of the
stellar population under scrutiny.  When properly calibrated, our
models with diffusion reproduce well within the errors the 
Hertzsprung-Russell diagram
of Hipparcos subdwarfs with empirically determined $T_{\rm eff}$ values
and high resolution spectroscopical [Fe/H] determinations.  

Since the observed surface abundances of subdwarfs are different from
the initial ones due to the effect of diffusion, while
the globular clusters stellar abundances are measured in Red Giants, which have
practically recovered their initial abundances after the dredge-up,
the isochrones to be employed for studying
globular clusters and Halo subdwarfs with the same
observational value of [Fe/H] are different and do not coincide.
This is at odds with the basic assumption of the MS-fitting 
technique for distance determinations.
However, the use of the rather large 
sample of Hipparcos lower MS subdwarfs with accurate parallaxes 
keeps at minimum the effect of these differences, for two reasons.
First, it is possible to use subdwarfs with observed [Fe/H] values
close to the cluster one; this minimizes the colour corrections
(which are derived from the isochrones)
needed to reduce all the subdwarfs to a mono-metallicity 
sequence having the same [Fe/H] than the cluster.
Second, one can employ objects
sufficiently faint so that the differences between the subdwarfs
and cluster MS with the same observed value of [Fe/H]
are small (they increase for increasing luminosity).
We find therefore that the distances
based on standard isochrones are basically unaltered
when diffusion is taken properly into account.

On the other hand, the absolute ages, the age dispersion, the
age-metallicity relation for Halo subdwarfs, as well as the value of the
helium enrichment ratio $\Delta Y/\Delta Z$ obtained from the
width of the empirical Halo subdwarfs MS, are all 
significantly modified when the properly calibrated isochrones with
diffusion are used.


\end{abstract}

\section{Introduction}

Atomic diffusion is a basic physical element transport mechanism
usually neglected in standard stellar models.  It is driven by
pressure gradients (or gravity), temperature gradients and composition
gradients. Gravity and temperature gradients tend to concentrate the
heavier elements toward the center of the star, while concentration
gradients oppose the above processes.  Overall diffusion acts very
slowly in stars, with time scales of the order of $10^{9}$ years, so
that the only evolutionary phases where diffusion is efficient are 
Main Sequence (but see also
Michaud, Vauclair \& Vauclair 1983 for a discussion about
the effect of diffusion in hot Horizontal Branch stars) and the 
White Dwarf cooling.

The occurrence of diffusion in the Sun has been recently demonstrated
by helioseismic studies (see, e.g, Chri\-stensen-Dalsgaard, Proffitt
\& Thompson 1993, Guenther, Kim \& Demarque 1996). Solar models
including this process can reproduce much better than standard models the
solar pulsation spectrum and the helioseismic values of helium surface
abundance and depth of the convective envelope.  Moreover, it seems
that neither turbulence nor other hydrodynamical mixing processes {\sl
substantially} reduce the full efficiency of element diffusion in the Sun,
otherwise the helioseismic constraints could not be satisfied 
(Richard et al. 1996).
Since
the Sun is a typical Main Sequence (MS) star whose structure closely
resembles the one of metal poor MS objects, it appears likely that
diffusion should also occur in MS globular cluster (GC) and field Halo
stars and be as efficient as in the Sun.
Very recently, Lebreton et al.~(1999) have shown 
that diffusion is necessary
for reproducing the effective temperatures of Hipparcos subdwarfs in the
metallicity range $-1.0 \leq$[Fe/H]$\leq -$0.3, belonging mainly to
the thick disk of the Galaxy. On the other hand 
the occurrence of a full efficiency of this process in Halo stars 
is still a matter of debate, since it appears to be unable to explain the
near constancy of the $^7{\rm Li}$ abundances in metal-poor  
stars with $T_{\rm eff}$ larger than $\sim$6000 K
(see, e.g., Vauclair \& Charbonnel 1998). As reviewed by 
Vandenberg, Bolte \& Stetson~(1996), turbulent mixing below the convective zone,
rotation, mass loss, have been proposed as additional processes able to partially
inhibit atomic diffusion; mass loss in particular (mass loss rates at the level
of $\approx 10^{-12}$--$10^{-13} M_{\odot} yr^{-1}$ would be necessary to reproduce 
the observations) is an interesting candidate
(Swenson 1995, Vauclair \& Charbonnel 1995), but there are no strong observational 
constraints at present.

Investigations dealing with the effect
of diffusion on Population II stars -- and the present work goes along
the same line -- have generally considered a full efficiency of this
process (but see also Proffitt \& Michaud 1991), and their
results can be regarded as an estimate of the upper limit of the
effect of atomic diffusion on the Halo stars evolution (another
non-conventional transport mechanism, radiative acceleration, does not
appear to appreciably affect the evolutionary properties of low mass
stars, at least in the range 1.1 $\leq M/M_{\odot}\leq$ 1.3, 
as investigated by Turcotte, Richer \& Michaud~1998).

Due to diffusion
(see, e.g., Castellani et al.~1997, Weiss \& Schlattl 1999) 
the stellar surface metallicity and helium content progressively decrease
during the MS phase -- due to their sinking below the
boundary of the convective envelope --, reaching a minimum around the
Turn-Off (TO) stage; then, since envelope convection deepens, a large
part of the metals and helium diffused toward the center are again engulfed
in the convective envelope, thus restoring the surface $Z$ 
(with $Z$ we indicate, as usual, the mass fraction of the metals) 
to {\sl almost} the initial value and $Y$ (helium mass fraction), 
after the first dredge up, 
to a value almost as high as for evolution without diffusion.
Along the Red Giant Branch (RGB)
phase diffusion is basically inefficient because of the much shorter
evolutionary time scales.  The net effect on the evolutionary tracks is
to have the MS (for a given stellar mass and initial chemical
composition) colder for fixed value of the luminosity, and a less
luminous and colder TO (which is reached earlier), with respect to
standard models.  The reason for this behaviour is that the inward
settling of helium raises the core molecular weight and the molecular
weight gradient between surface and center of the star. This increases
the stellar radius and the rate of energy generation in the
center. The metal diffusion only partially counterbalances this effect
by decreasing the opacity in the envelope and increasing the central
CNO abundance.

With respect to a standard isochrone of given initial metallicity and
reference TO luminosity, isochrones computed accounting for the
diffusion of helium and metals give an age lower by $\simeq$1 Gyr,
if the same initial metallicity is used.
As far as the RGB evolution is concerned, the location of
the RGB in the Hertzsprung-Russell (HR) diagram is basically unchanged
with respect to standard models, and the level of the Horizontal
Branch is also almost negligibly affected
(Castellani et al.~1997).

Several papers have been published about the influence of atomic
diffusion in old stars (see, e.g., Proffitt \& Vandenberg 1991,
Chaboyer, Sarajedini \& Demarque 1992, D'Antona, Caloi \& Mazzitelli
1997, Castellani et al. 1997, Cassisi et al. 1998,
Castellani \& Degl'Innocenti 1999), with
the main goal of studying the influence on GC ages. The
approach usually followed is to compute models with diffusion
for a certain set of initial metallicities, and then to
compare with the observational Colour-Magnitude-Diagram (CMD) of a
given GC the isochrones whose initial metal content matches the
spectroscopical GC one.  Since the chemical composition of a GC is
determined by means of observations of its RGB stars (see, e.g.,
Carretta \& Gratton 1997), the spectroscopical GC metallicity truly
reflects the initial one.

A very different situation holds for field MS subdwarfs, a point
recently raised by Morel \& Baglin (1999 - hereinafter MB99).  
The spectroscopical subdwarfs metallicity is not the original
one, because diffusion along the MS decreases the envelope metallicity
(and helium content). When comparing theoretical diffusive isochrones
with subdwarfs, one must compute models with suitable initial chemical
abundances which produce, at the subdwarf age, its observed surface
metallicity. This means that subdwarfs models must be computed using a
larger initial $Z$ with respect to the observed one, the exact value
depending on the star age.  This occurrence has potentially very
important implications for example for MS-fitting distances and
subdwarfs GC ages, since it implies that the MS (and TO) of subdwarfs
and GC sharing the same observational value of the metallicity are not
coincident.

In this paper we present models for the MS phase of low mass 
metal-poor stars, accounting for
atomic diffusion (Sect. 2); we have considered the full efficiency
of this process, as in the Sun. We will then analyze for the
first time in a consistent way the effect of diffusion on three
important quantities for cosmological and Galactic evolution models,
namely GC distances derived by means of the MS-fitting technique
(Sect. 3), age of field subdwarfs
(Sect. 4), and the helium enrichment ratio $\Delta Y/\Delta Z$ 
estimated from the width of the local subdwarfs MS
(Sect. 5). A summary follows in the final section.
In view of the previously discussed possibility that 
the efficiency of diffusion could be somewhat reduced in Population II stars,
our results may be viewed as upper limits.

\section[]{The models}

The stellar evolution computations have been performed using the same
code and the same input physics as in Salaris \& Weiss (1998 -
hereinafter SW98). We just recall that we have used the opacities by
Iglesias \& Rogers (1996) and Alexander \& Ferguson~(1994) for an
$\alpha$-enhanced heavy elements mixture
($\langle[\alpha/{\rm Fe}]\rangle$=0.4; the details of the distribution
are given in SW98), and that the mixing length 
calibration allows to reproduce the RGB effective
temperatures of a selected sample of
GC in the ($M_{\rm bol}, T_{\rm eff}$) plane, as derived by Frogel, Persson \&
Cohen~(1983), which is the observational constraint for Halo stars.  
In case of models with diffusion, since the RGB
position is almost unchanged with respect to standard isochrones, the
same mixing length value satisfies the metal-poor RGB observational
constraint. Therefore, we did not change the mixing length calibration
with respect to the case of standard isochrones.  This permits also to
clearly show the influence of diffusion on the stellar models without
any contribution from the variation of other parameters.
We stress also that large part of our results does not depend
on the mixing length calibration. 

For the
present calculations we have included the diffusion of
helium and heavy elements following Thoul, Bahcall \& Loeb~(1994);
their formalism and the input physics used in our models have been
already successfully tested on the Sun
(see, e.g., Ciacio, Degl'Innocenti \& Ricci~1997).
The variations of the abundances of H, He, C, N, O and Fe are followed
all along the structures; all other elements are assumed to diffuse in
the same way as fully-ionized iron (see, e.g., Thoul et al.~1994
for a comparison among different diffusion formalisms). The local
changes of metal abundance are taken into account
also in the opacity computation, by calculating the actual global
metallicity at each mesh point, and then interpolating among tables with
different $Z$. In this way one does not take into account the {\sl
small} changes in the metal distribution due to differences in the
diffusion velocities of C,N,O and Fe; however, for the mass range we
are dealing with ($M\leq$1.0 $M_{\odot}$) the differences are small
and our procedure does not introduce any significant error in the
opacity (see, e.g., the detailed discussion in Turcotte et
al.~1998).

We have computed a set of MS models (and isochrones) with diffusion
and initial metallicities [Fe/H]=$-2.3$, $-2.0$, $-1.7$, $-1.3$,
$-1.0$, $-0.7$, $-0.6$
($Y_{0}$=0.23 and $\Delta Y/\Delta Z$=3 as in SW98). In
addition we have also computed a set of isochrones for 8 and 12 Gyr
which displays as {\sl actual} surface metallicity the set of values
previously given (we will call them `calibrated' diffusive isochrones,
following the nomenclature by MB99).  When computing the latter
isochrones we had to ensure that, for the selected ages, all the
different evolving masses showed the prescribed surface
metallicity. This means that we had to employ an iterative procedure
for finding the exact value of the initial metallicity (larger than
the final one) to be used for each case (we kept fixed $\Delta
Y/\Delta Z$=3 when deriving the initial helium abundance).

In Figure 1 we show a comparison in the HR diagram among the
sets of isochrones computed, and the standard ones by SW98, with $t=$8
and 12 Gyr and [Fe/H]=$-$2.3 and $-1.0$.  For the standard
(solid lines) and `calibrated' diffusive ones (dotted lines -
hereinafter C) the labelled value of [Fe/H] represents the actual
surface metallicity, which is constant along the isochrone, while for
the non calibrated isochrones with diffusion (dashed line -
hereinafter D) it represents only the initial surface metallicity.

\begin{figure}
\psfig{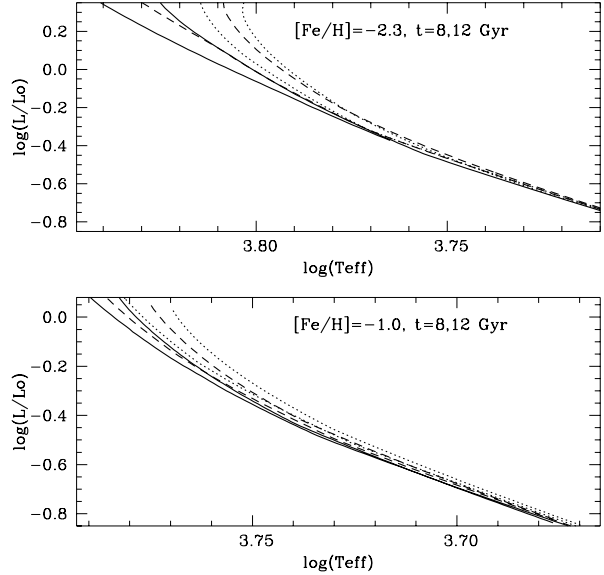}
\caption[]{HR diagram of standard (solid line), D (see text - dashed
line) and C isochrones (see text - dotted line) with $t=$8,12 Gyr,
[Fe/H]=$-$2.3 (upper panel) and [Fe/H]=$-$1.0 (bottom panel).}
\end{figure}

Our results closely
resemble the results by MB99; specifically, for a fixed age and
luminosity the standard isochrones are hotter than the D and C
ones. The C isochrones are the reddest among the three sets. The
$T_{\rm eff}$ difference at fixed luminosity among the three sets of
isochrones increases with increasing metallicity.  As in MB99 we find
that the shift toward higher $T_{\rm eff}$ of the C isochrones with
respect to the standard ones increases for increasing luminosities;
this changes slightly the slope of the MS, making it more vertical for
the C isochrones. At [Fe/H]=$-$2.3 the C and D isochrones are almost coincident 
all along the lower MS.
Table 1 displays a quantitative evaluation of the $T_{\rm eff}$ difference, 
at selected luminosities along the lower MS, between standard and C isochrones 
with $t$=8 and 12 Gyr and metallicities [Fe/H]=$-$1.0, $-$1.7, $-$2.3.

\begin{figure}
\psfig{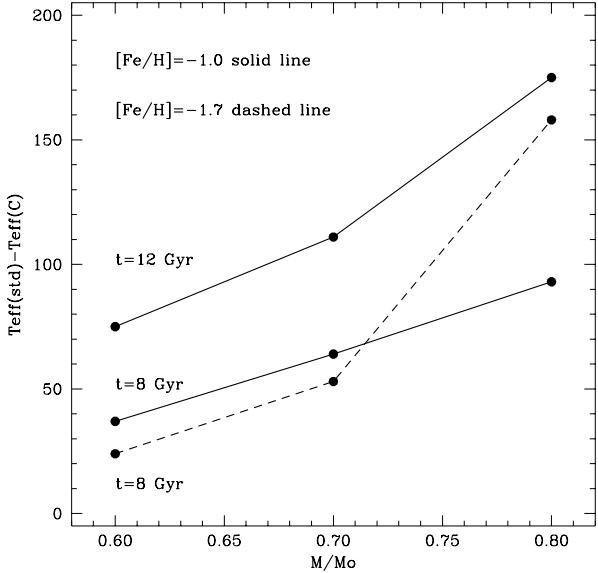}
\caption[]{Difference $\Delta T_{\rm eff}$ (K) between standard and
calibrated diffusive C isochrones for three selected values of the
evolving mass, with metallicities [Fe/H]=$-1.0$ ($t=$8 and 12 Gyr) and
$-1.7$ ($t=$8 Gyr). The crossing between the [Fe/H]=$-1.7$ models and the
more metal rich ones is due to evolutionary effects.}
\end{figure}

Figure 2 shows the difference $\Delta T_{\rm eff}$ at fixed age between
standard and C isochrones for three selected values of the evolving
mass along the MS, metallicities [Fe/H]=$-$1.0 and $-$1.7, $t=$8 and 12 Gyr.  
For the [Fe/H]=$-1.7$ isochrones only data with $t=$8 Gyr are shown, since stars
with M=0.8$M_{\odot}$ are evolved off the MS at $t=$12 Gyr. The
displayed data are comparable with analogous quantities in Fig. 6 of
MB99; our results appear consistent with MB99, when taking into
account that their $\Delta T_{\rm eff}$ values correspond to an age of 10
Gyr.

\begin{figure}
\psfig{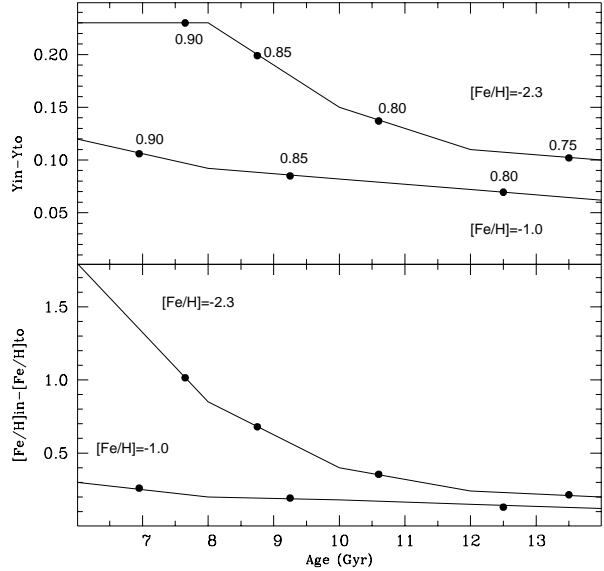}
\caption[]{He and metal depletion at the TO
(difference between the initial and the TO values) 
as a function of the age for selected D isochrones with the displayed 
initial metallicities. Filled circles show selected TO masses 
indicated in units of $M_{\odot}$ (upper panel).}
\end{figure}

In Figure 3 the differences $\Delta$[Fe/H]
($\Delta Y$) between the initial metallicity (helium content)
and the TO surface values of D isochrones
are displayed for two selected initial metallicities and various ages.
As a general trend, $\Delta$[Fe/H] and $\Delta Y$ 
increase for decreasing age (in the age range we are dealing with)
and for decreasing initial metallicity (at fixed age).
In the age range 6-14 Gyr $\Delta$[Fe/H] is generally between
0.1 and 1.0 -- apart for the most metal poor isochrones 
which show a larger metal depletion
($\Delta$[Fe/H]$\simeq$1.8 at $t$=6 Gyr for an initial [Fe/H]=$-$2.3) -- , 
while $\Delta Y$ ranges between $\sim$0.05 and $\sim$0.2;
the surface $Y$ is practically zero for the
[Fe/H]=$-$2.3 isochrones when $t$ is less than about 8 Gyr.

\begin{table*}
\caption[]{$T_{\rm eff}$ difference (in K) between standard and C isochrones 
($T_{\rm eff}$(standard)-$T_{\rm eff}$(C)) for
three selected metallicities ([Fe/H]=$-$2.3, $-$1.7, $-$1.0)
and three luminosities along the lower MS.}
\begin{tabular}{cccccccc} \hline
 & & $t$=8 Gyr & & & & $t$=12 Gyr & \\
$\log(L/L_{\odot})$&$-2.3$&$-1.7$&$-1.0$& &$-2.3$&$-1.7$&$-1.0$\\   
\hline
-0.4 & 37 & 48 & 64 & &47 & 65 & 90\\
-0.6 & 29 & 40 & 52 & &37 & 54 & 76\\
-0.8 & 20 & 30 & 41 & &28 & 45 & 63\\
 \hline
\end{tabular}
\end{table*}

This increase of the TO surface metallicity (and helium) depletion with
decreasing age could appear at first surprising, since diffusion has
less time to work when age is lower; however, one must also take into
account that convective envelopes are progressively thinner for stars
populating the TO at decreasing age, thus increasing the rate of
depletion of helium and metals (notice that more metal poor models
have thinner surface convective regions). It is the competition between
these two factors which determines the final trend of the TO surface
abundances with time.

In Fig. 4 we compare our C isochrones (bolometric
luminosities were transformed to $M_{\rm V}$ using the bolometric
corrections described in Weiss \& Salaris 1999; we just recall that,
after the necessary calibration of the zero point to reproduce the solar $M_{\rm V}$
value, our adopted bolometric corrections from Buser \& Kurucz 1978, 1992
agree quite well with the empirical 
determinations by Alonso, Arribas \& Martinez-Roger 1996a) 
with a sample of metal-poor subdwarfs with
accurate Hipparcos parallaxes ($\sigma(\pi)/\pi <0.12$) listed by
Carretta et al.~(1999), and $T_{\rm eff}$ derived from the Infrared Flux
Method (IRFM -- Alonso, Arribas \& Martinez-Roger 1996b).  
The goal is to check
if isochrones with full efficiency of diffusion 
are compatible with the HR diagram of field subdwarfs
with accurate parallaxes and empirical $T_{\rm eff}$ determinations
(therefore eliminating the influence of the adopted colour-transformations).

Lebreton et al.~(1999) recently performed this kind of
comparison for subdwarfs with metallicities in the range $-$1.0$<$[Fe/H]$<$0.3,
and found that for $-$1.0$<$[Fe/H]$<-$0.5 the inclusion of the  
full efficiency of diffusion is necessary 
for reproducing observational data. We extend their analysis by studying 
the case of lower metallicities. Our adopted subdwarfs
spectroscopic [Fe/H] values come from Carretta et al. (1999), and are
homogeneous with the Carretta \& Gratton (1997) metallicity scale for
GC; even if the metallicities adopted by Alonso et al.~(1996b) for
applying the IRFM method are different (generally lower by $\approx$
0.2), the sensitivity of the derived $T_{\rm eff}$ to the input
metallicity is so low (Alonso et al.~1996b) that no appreciable
inconsistency is introduced by our choice of the [Fe/H] scale.  
For two stars we found differences as high as $\approx$1.0 between the 
metallicity used by  Alonso et al.~(1996b) and the 
Carretta et al.~(1999), and we did not consider them.
The [Fe/H] values for the subdwarfs displayed in the pictures are in the
ranges, respectively, between $-$1.55 and $-$1.69  (top panel), 
$-$1.24 and
$-$1.28  (middle panel), $-$0.98 and$-$1.02 (bottom panel).  The
error bar on [Fe/H] is typically by 0.10-0.15, while the errors on
$T_{\rm eff}$ and $M_{\rm V}$ are shown in the figure.

\begin{figure}
\psfig{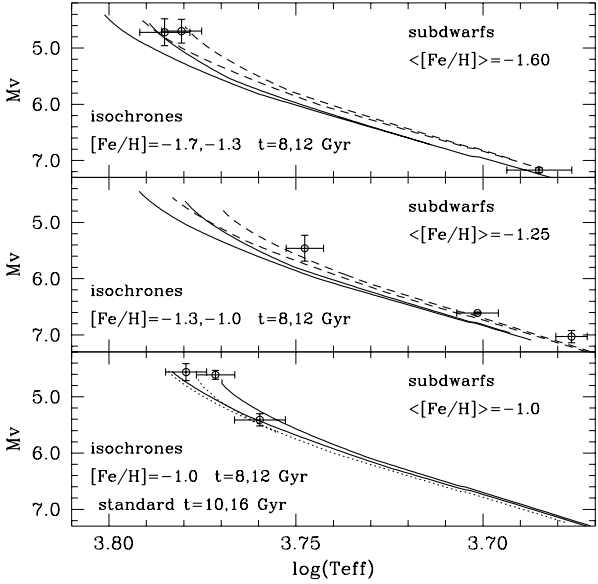}
\caption[]{$M_{\rm V}$-$T_{\rm eff}$ diagram of C isochrones and a sample of
subdwarfs with accurate Hipparcos parallaxes and IRFM effective
temperatures. The average subdwarfs metallicity is given in the 
labels at the top right corner of the three panels; isochrones
metallicities and ages are shown at the left bottom corner of each panel.
Subdwarfs are represented by circles; error bars on their values
of $M_{V}$ and log($T_{\rm eff}$) are also displayed.
In the upper and middle panel solid lines correspond to the 
C isochrones with the lower metallicity between the given values. 
In the bottom panel C isochrones are represented with
solid lines, while the standard isochrones 
with [Fe/H]=$-$1.0 and t=10,16 Gyr
are also displayed (dotted lines) for comparisons.
}
\end{figure}

The displayed MS isochrones have ages equal to 8 and 12 Gyr
respectively, corresponding to approximately the upper and lower limit
of the ages determined by SW98 for a large sample of Galactic GC.  As
it is evident from the figure, the agreement between observations and
theoretical models is satisfactory. Our calibrated diffusive C
isochrones (which are the correct ones to be compared with field
subdwarfs of a given metallicity) reproduce satisfactorily the
observations. The best agreement is for subdwarfs with average
metallicity [Fe/H]=$-$1.0 and $-$1.6; at the intermediate metallicity
$-1.25$ the models appear to be slightly too hot, but when taking into
account the error bar on the temperature and metallicity of the
subdwarfs the discrepancy does not appear to be significant.
In the case of [Fe/H]=$-$1.0 standard isochrones are also shown;
it is clear that in spite of the good agreement between C models and data,
one cannot use these results as a definitive proof that diffusion is fully
efficient in Halo subdwarfs. Standard isochrones can also 
reasonably reproduce the observational data (even if for different ages), 
at least given the present sample of objects and 
observational uncertainties on $M_{\rm V}$ and $T_{\rm eff}$.

Regarding this last point one should notice that
Cayrel et al (1997) have shown how the MS location of standard isochrones
appears to be too hot by $\approx$120-140 K
in comparison with a sample of metal poor subdwarfs with accurate Hipparcos parallaxes,
when using the empirical $T_{\rm eff}$ determinations by Alonso et al.~(1996b), 
and [Fe/H] values from the compilations by Cayrel de Strobel et al. (1997). This difference 
with respect to our conclusion that at [Fe/H]=$-$1.0 standard isochrones are still able to reproduce 
within the errors the position of local subdwarfs, is due mainly to the different [Fe/H] scale 
we have employed. We used the Carretta et al (1999) metallicities which are on a scale
homogeneous with the GC metallicities we will use in the next section. For the stars with
[Fe/H]$\sim -$1.0 our adopted [Fe/H] values are about 0.2-0.3 larger than the corresponding 
values given by Cayrel de Strobel et al. (1997 - we averaged the high S/N data); 
these larger metallicities reduce the discrepancy between standard isochrones and observations.

\section[]{Atomic diffusion and MS-fitting distances}

The MS-fitting technique is the `classical' method to derive distances
to GC (see, e.g., Sandage 1970).  The basic idea is very simple.
Suppose that precise parallaxes of neighbouring subdwarfs are
available; for a given GC metallicity $Z_{\rm cl}$, it is possible to
construct an empirical template MS by considering subdwarfs with
metallicity $Z_{\rm sbdw}$ close to $Z_{\rm cl}$ and applying to their
colours small shifts (obtained using the derivative
$\Delta(colour)_{\rm MS}/\Delta$[Fe/H] as derived from theoretical
isochrones) for reducing them to a mono-metallicity sequence with
$Z = Z_{\rm cl}$.  The fit of this empirical MS to the observed GC one
(reddening-corrected) provides the cluster distance modulus.

The main underlying assumption is that the MS of subdwarfs with a
certain value of [Fe/H] is {\sl coincident} with the MS of GC with the
same metallicity.  If atomic diffusion is at work in Halo stars, the
underlying assumption of this technique is no longer rigorously
satisfied. The point is that (as previously discussed) the
spectroscopical metallicity of a GC is determined from its RGB stars;
this metallicity is very close to the primordial GC chemical composition, but
is not (due to the effect of diffusion) the MS one. Therefore, when
fitting the local subdwarfs MS to the MS of a GC with the same observed
metallicity, one is introducing an error in the derived distance
modulus.

The release of the Hipparcos catalogue has enlarged the number of
metal poor subdwarfs with precise parallaxes which can be now used for
applying this technique to the Galactic GC, and several authors
(see, e.g, Reid 1997, Gratton et al.~1997, Chaboyer et al.~1998,
Carretta et al.~1999) have recently derived distances (and ages) of GC
using the MS-fitting and subdwarfs with Hipparcos parallaxes.  In
particular, Carretta et al.~(1999) have carefully analyzed the total
error budget associated with the MS-fitting, but the effect of atomic
diffusion in subdwarfs is nowhere mentioned.

\begin{figure}
\psfig{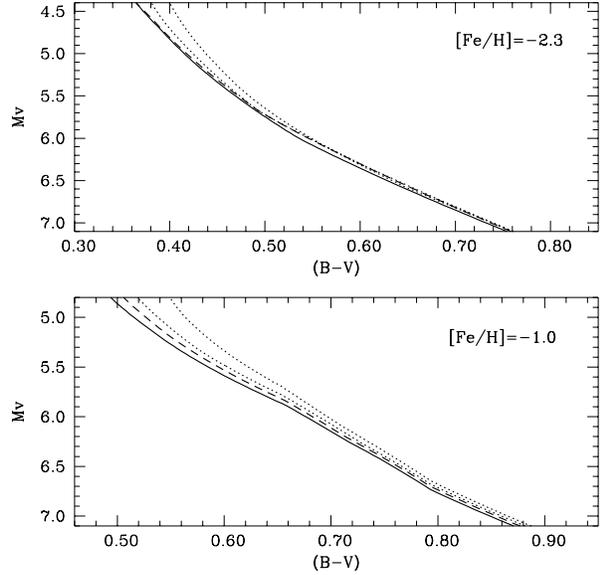}
\caption[]{Colour Magnitude Diagram of standard (solid line) and D
(dashed line) isochrones with $t=$8 Gyr and the displayed metallicities;
also plotted are C isochrones (dotted lines) for the same
metallicities but $t=$8 and 12 Gyr.}
\end{figure}

We display in Fig. 5 standard, C and D isochrones transformed to the
$M_{\rm V}-(B-V)$ plane according to the transformations described in 
Weiss \& Salaris (1999), for
[Fe/H]=$-2.3$ and $-1.0$.  In the case of D isochrones, when computing
the bolometric corrections (already used in the previous section) and
the $T_{\rm eff}$-colour conversion, we have taken into account the fact
that the surface [Fe/H] value is not constant along the MS (it is
actually decreasing).  It must be noticed that the surface helium
content for stars around the TO of the isochrones with diffusion is
always much lower (generally $Y$ is in the range $\approx$0.1-0.2 for
ages between 8 and 12 Gyr, as discussed in the previous section) 
than the helium content used in the model
atmospheres producing the adopted colour transformations.  However,
this does not introduce a substantial systematic error since,
according to Carney (1981), the He abundance does not appreciably
affect the flux distribution at temperatures appropriate to GC dwarfs
and subgiants.

The behaviour of the isochrones in the observational $M_{\rm V}-(B-V)$ plane
closely follows the results in the theoretical HR diagram.
Standard isochrones (solid line) are
systematically bluer than the diffusive ones.  The C isochrones are
the reddest ones, and they are progressively redder than the standard
or the D ones for increasing age (the effect is stronger for large
metallicities). In Fig. 6 the lower MS for [Fe/H]=$-1.0$ and $t=$8,12
Gyr is shown; when $M_{\rm V}\leq$6 standard isochrones are unaffected by age
(a fact that is well known), while D isochrones are insensitive to
the age only for $M_{\rm V}\leq$7 and C isochrones are always affected by
the age, at least down to $M_{\rm V}$=7.3 (because of the 
metal decrease with time due to diffusion).

\begin{figure}
\psfig{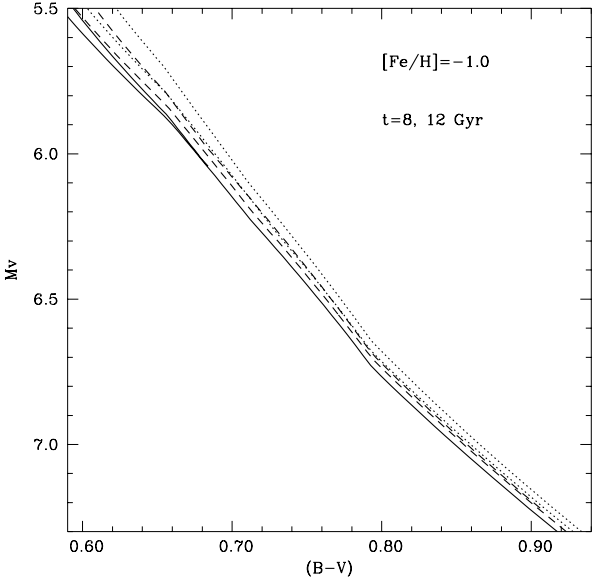}
\caption[]{Colour Magnitude Diagram of the low MS of standard (solid
line), D (dashed line) and C isochrones (dotted lines) with $t=$8 and 12
Gyr and [Fe/H]=$-1.0$.}
\end{figure}

The MS of a GC with an observed RGB metallicity
[Fe/H]=$-1.0$ is given by the D isochrones in Fig. 6, while the MS
of local subdwarfs with the same observed metallicity is given by the
C isochrones. As it is evident {\sl the two MS are generally not coincident}.

\begin{table*}
\caption[]{MS-fitting distance moduli ($(m-M)_{V}$) of selected clusters.}
\begin{tabular}{clllll} \hline
& & \multispan2{\hfil$t_{\rm cl}$=8Gyr\hfil} &\multispan2{\hfil$t_{\rm
cl}$=12Gyr\hfil} \\
Cluster&standard&$t_{\rm sbdw}$=8 Gyr & $t_{\rm sbdw}$=12 Gyr  & $t_{\rm sbdw}$=8 Gyr & $t_{\rm sbdw}$=12 Gyr \\
\hline
M92    & 14.76$\pm$0.04 & 14.75$\pm$0.04 & 14.74$\pm$0.04 & 14.75$\pm$0.04 & 14.74$\pm$0.04 \\
M5     & 14.59$\pm$0.03 & 14.56$\pm$0.03 & 14.53$\pm$0.03 & 14.58$\pm$0.03 & 14.54$\pm$0.03 \\
NGC288 & 14.93$\pm$0.03 & 14.90$\pm$0.03 & 14.88$\pm$0.03 & 14.92$\pm$0.03 & 14.89$\pm$0.03 \\
47 Tuc & 13.58$\pm$0.04 & 13.59$\pm$0.04 & 13.56$\pm$0.04 & 13.63$\pm$0.04 & 13.59$\pm$0.04 \\
\hline
\end{tabular}
\end{table*}

Another difference with respect to the standard case (and another {\sl
potential} source of systematic errors on the actual MS-fitting
distances) is the value of the derivative $\Delta (B-V)_{\rm MS}$/$\Delta$
[Fe/H]. This is the only information needed from theoretical
isochrones to be employed in the MS-fitting technique; it is used for
shifting the subdwarfs to a mono-metallicity sequence corresponding to
the observed cluster [Fe/H]. Since the difference in colour between
the standard MS and the C MS depends on the metallicity (see Fig. 5)
this will have an impact on $\Delta(B-V)_{\rm MS}/\Delta$[Fe/H] for the
subdwarfs. 

Are these differences large enough to affect substantially 
the MS-fitting GC distances? 
It depends on the subdwarf sample. To explain this point
let's consider, as an example, 
subdwarfs with [Fe/H]$\simeq -$1.3 and $M_{\rm V}\simeq 6$,
which hypothetically have to be employed for deriving the distance to a GC with 
[Fe/H]$\simeq -$0.7.  The variation of
$\Delta(B-V)_{\rm MS}/\Delta$[Fe/H] due to diffusion causes 
a shift of the empirical subdwarfs MS at the cluster metallicity
by $\Delta (B-V)\simeq$+0.02 with respect to the standard case.
This, by itself, would induce a GC distance modulus larger by $\simeq$0.1 mag, since
the MS slope $\Delta M_{\rm V}/\Delta (B-V)$ is equal to about 5.5.
However, one must correct for the vertical $M_{\rm V}$ difference
between subdwarfs and GC MS, which tends to reduce
the derived distance modulus by $\approx$0.05--0.08 mag in the age range 8--12 Gyr.
The final combined effect is to have distances unchanged or increased at most by 
0.05 mag with respect to the standard case.
However, in the hypothesis that for determining the MS-fitting distance to a GC with 
[Fe/H]$\simeq -$1.3 one can use only subdwarfs with [Fe/H]$\simeq -$0.7, 
the situation is quite different, since the use of the diffusive C isochrones would
cause a decrease of the distance modulus by $\approx$0.10-0.13 mag.

In the following we will study the effect of diffusion on the 
MS-fitting distances obtained using subdwarfs with accurate Hipparcos parallaxes.
We have considered, as a test (the results are summarised in Table 2),
four clusters included in the analysis by Gratton et al.~(1997), namely M92
([Fe/H]$\simeq -2.15$), M5 and NGC288 ([Fe/H]$\simeq -1.1$), 47Tuc 
([Fe/H]$\simeq -0.7$).  The
subdwarfs $M_{\rm V}$, $(B-V)_{0}$ and [Fe/H] values come from Table 2 of
Gratton et al.~(1997); the clusters reddenings and metallicities are
from the quoted paper, as well as the observational clusters MS lines.
For each cluster we have considered only {\sl bona fide}
single stars fainter than $V$=6 (to avoid evolutionary effects for the
standard isochrones, as well as the influence of the mixing-length
calibration), with $\sigma(\pi)/\pi <0.12$ and 
in the same metallicity range as in Gratton et al.~(1997).

In the case of the standard models by SW98 we recover basically the
same distances by Gratton et al~(1997), whose results were obtained by
using a value for $\Delta (B-V)_{\rm MS}/\Delta$ [Fe/H] derived from
different isochrones, and considering subdwarfs also in the range
5.5$<M_{\rm V}<$6.0.  
When deriving the MS-fitting distances taking into account diffusion, 
we have (as outlined in the previous example) corrected the subdwarfs colours by using 
the $\Delta (B-V)_{\rm MS}/\Delta$ [Fe/H] values derived from the C models, 
and we have also accounted for the difference in brightness 
at fixed colour between
the subdwarfs MS (C isochrones) and the clusters one (D isochrones).
Since there are {\sl small} evolutionary effects for the D isochrones
(representing the GC) even when  6$\leq M_{\rm V}\leq$7,
we have taken into account 4
different possibilities.  In the
first two cases we have assumed for the clusters age
$t_{\rm cl}$=8 Gyr with subdwarfs ages $t_{\rm sbdw}$=8 and
12 Gyr, and in the second two cases we considered $t_{\rm cl}$=12
Gyr and again $t_{\rm sbdw}$=8 and 12 Gyr.

As it is clear from Table 2, there are no appreciable modifications to
the distance moduli derived from standard isochrones. The differences
with respect to the standard case are small and generally within the {\sl small}
formal error bars associated to the fit (the error bar takes into
account {\it only} the error on the fit due to the uncertainties on the
subdwarfs $M_{\rm V}$ and $(B-V)$).  This is a quite important point, since it
confirms the robustness of the published Hipparcos MS-fitting distances
which did not take into account the effect of atomic diffusion on GC
and field subdwarfs evolution.

The reason for this occurrence is that -- thanks to the Hipparcos
results -- the sample of lower MS metal poor subdwarfs with accurate
parallaxes has substantially increased with respect to the recent past. In
performing the MS fitting we have used objects whose metallicity is
close to the actual GC metallicity; in this case, as it is evident,
the colour correction to be applied to the subdwarfs is small, and
even the occurrence of a
sizeable change of $\Delta (B-V)_{\rm MS}/\Delta$ [Fe/H] does not
modify appreciably the final distance.  Moreover, the subdwarfs are
all sufficiently faint so that the difference between the GC (D
isochrones) and subdwarfs (C isochrones) MS is 
generally kept at the lowest possible  
value (this difference generally increases for increasing luminosity).

In conclusion, the effect of diffusion on the two main distance
determination methods for GC stars, namely MS-fitting and HB fitting,
is practically negligible, since also the HB luminosities are
negligibly influenced by diffusion. The final effect on the GC age
estimates is therefore just a reduction by about 1 Gyr due to the
change of the TO brightness.

\section{Atomic diffusion and subdwarfs ages}

The age of field subdwarfs is, together with the age of stellar
clusters, an important piece of information needed for understanding
time scales and formation mechanism of the Galaxy.  As repeatedly
stressed before, the
isochrones used for studying field subdwarfs of a certain surface
metallicity are different from the ones to be employed when dealing
with GC with the same observational value of [Fe/H].  This
is clearly at odds with the usual procedure, when using standard
isochrones, to use the same models for field and GC MS stars.

In Fig. 7 (upper panel) we display, as an example, standard and C
isochrones with [Fe/H]=$-$2.3 and selected ages ($t=$8,12,13 Gyr in the
case of standard models, while for C isochrones $t=$8,12 Gyr). It
appears clear how the influence of diffusion is extremely relevant for
the derived subdwarfs ages.  The TO absolute $V$ brightness for a standard
isochrone with $t=$13 Gyr is coincident with the TO brightness of a C
isochrone of {\sl only} 8 Gyr; both the TO $M_{\rm V}$ and $(B-V)$ differences
corresponding to an age between 8 and 12 Gyr are strongly reduced (by
$\simeq$50\%) when passing from standard to C isochrones.  This
undoubtedly causes a strong reduction in the derived subdwarfs age,
and has an important effect also on the determination of the age
dispersion.

The reason for such a big difference with respect to standard
calculations is that the C isochrone is basically an isochrone of much
larger initial metal abundance than the standard one (by how much
larger depends on the age) and with diffusion. It is well known that
for a fixed age the effect of diffusion {\sl at a fixed initial
metallicity} is to decrease the TO luminosity (and colour); in the
case of C isochrones, there is in addition the effect of the larger
initial abundance which further lowers the TO luminosity (and
colour); the amount of the cumulative effect depends on the
selected age and initial metallicity.

In the lower panel of Fig. 7 a comparison between the TO region of C
and D isochrones with [Fe/H]=$-$2.3 and $t=$8,12 Gyr is shown. The use
of the D isochrones (suitable for GC) for deriving subdwarfs ages does
not introduce a too large error for $t=$12 Gyr, while the difference
with respect to C isochrones is very large for $t=$8 Gyr. This is due
to the fact that, due to the lower age, the TO-mass is higher
and therefore the depth of the convective envelope is smaller in TO stars,
with the consequent larger depletion  
(and increase of the initial metallicity for the calibrated models)
of the metal and helium abundances due to diffusion.

\begin{figure}
\psfig{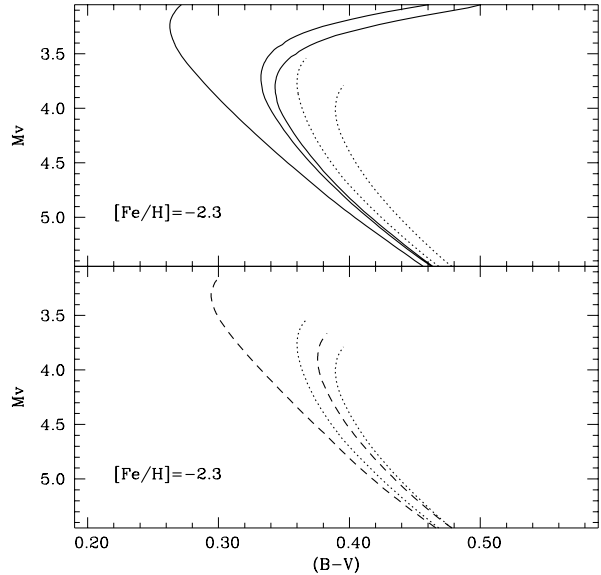}
\caption[]{{\sl Upper panel:} Colour Magnitude Diagram of the TO
region of standard (solid lines) and C isochrones (dotted lines) for
the displayed value of [Fe/H]; the ages are 8,12,13 Gyr for the
standard isochrones, 8 and 12 Gyr for the C ones.  {\sl Lower panel:}
as in the upper panel, but for D (dashed lines) and C (dotted lines)
isochrones with $t=$8,12 Gyr.}
\end{figure}

Diffusion has further important consequences when trying to determine
the age-metallicity relation for old Halo subdwarfs. We have already
seen that the use of diffusive C isochrones in place of standard ones
strongly reduces the derived subdwarfs ages.  Once the age of a sample
of subdwarfs is obtained, one can study the age-metallicity relation
for deriving information about Galactic formation mechanism and
time scales. Of course the relevant quantity is the relation between
the age of the stars and the metallicity from which they formed,
which, in case of diffusion, is different from the actual one. The
difference between initial and surface [Fe/H] at the TO is always
within 0.1-0.3 for ages of about 10 Gyr or larger.  But if one
derives subdwarfs ages of the order of 6-8 Gyr
(using the appropriate C isochrones), the observed [Fe/H] of
TO stars must be increased by $\approx$0.2-1.0 
(somewhat larger corrections are found for the most metal poor models
with [Fe/H]=$-$2.3) for obtaining the
initial value. This has to be taken into account in the analysis.

It is interesting to notice that if, for example, TO very metal poor
subdwarfs with [Fe/H] around $-$3.0 (a sample of them can be found
in Schuster et al.~1996) are found to be relatively young ($t=$8 Gyr)
when employing the appropriate $\alpha$-enhanced diffusive C
isochrones, their initial metallicity had to be [Fe/H]$\simeq -$2.2,
very similar to the initial metallicity of the most metal poor GC.

By considering the 
fact that the [Fe/H] depletion is larger at lower ages
(at least for ages larger than $\approx$6 Gyr), it is interesting to notice
that if one finds in a given observed metallicity range
that metal poor subdwarfs are younger than metal rich ones, 
this could correspond -- for a particular combination of ages
and width of the metallicity range -- to an age spread  
at a {\sl constant} value of the initial metallicity.

\section{Atomic diffusion and the value of $\Delta Y/\Delta Z$}

The ratio $\Delta Y/\Delta Z$ of helium supplied to the interstellar
medium by stars relative to their supply of heavy elements is an
important quantity to test theoretical stellar yields and for deriving
the slope of the relation between helium and oxygen in extra-galactic
H{\sc ii} regions, a fundamental ingredient for determining the
primordial helium abundance.

One possible approach to the determination of this quantity is the
study of the MS width of local subdwarfs. Since changes of the initial
values of $Y$ and $Z$
push the MS in opposite directions (increasing $Y$ makes the MS bluer,
while increasing $Z$ makes it redder), the width of the local
subdwarfs MS for a fixed metallicity range is a function of the $\Delta
Y/\Delta Z$ ratio in the interstellar medium. 
One can therefore consider two $\Delta Y/\Delta Z$ indicators:
either the vertical (usually in $M_{\rm bol}$) width at a fixed value
of $T_{\rm eff}$,
or the horizontal width (in $(B-V)$ or log$(T_{\rm eff}$)) at a fixed
value of $M_{\rm V}$
(see, e.g., Castellani, Degl'Innocenti \& Marconi 1999 
and references therein).
Usually the lower MS
(corresponding to subdwarfs  with $M_{\rm V}>$5.5-6.0) is used for
the analysis to avoid (as in the MS-fitting technique) evolutionary
effects and the influence of the mixing length calibration.

As it has been shown before, 
one of the effects of atomic diffusion on MS subdwarfs is to increase
the MS width for a fixed metallicity interval and assumed initial $\Delta
Y/\Delta Z$ value.  This is due to the fact that the colour difference
between the diffusive C isochrones and the standard ones is
metallicity dependent, and is larger at larger metallicities.

As an example, we have considered  
a value for subdwarfs effective temperature log($T_{\rm eff}$)=3.70
(corresponding to $M_{\rm V} >$6);
we then computed the MS $\Delta M_{\rm bol}$ broadening due to diffusion,  
in the interval between [Fe/H]=$-$2.3 and $-$0.7 -- a metallicity range typical of 
Halo subdwarfs -- and for subdwarfs ages equal to 8 and 12 Gyr, by means of comparisons
with the SW98 models.
As expected,  $\Delta M_{\rm bol}$  results to be larger for C isochrones
with respect to standard ones, the exact value depending on the subdwarfs age
since the entire MS location of C isochrones does depend on age; this means that 
standard isochrones underestimate $\Delta Y/\Delta Z$ with
respect to the calibrated diffusive ones.

The amount of this systematic difference was
derived by computing additional C isochrones and varying the initial 
$\Delta Y/\Delta Z$ ratio in the range between 1 and 5. 
We found that C isochrones (in the explored $\Delta Y/\Delta Z$ range) lead to
initial $\Delta Y/\Delta Z$ ratios larger by $\approx$1-2,  
the exact amount depending on the subdwarfs ages.
Moreover, we found that the dependence on the initial helium abundance of 
the values of $M_{\rm bol}$ at a fixed log$(T_{\rm eff})$  
along the lower MS, is in broad agreement with the results from 
standard models by Castellani et al.~(1999).

\section{Summary}

We have analyzed the influence of heavy elements and helium diffusion
on the MS of metal poor low mass stars in connection with the
determination of GC distances via MS-fitting technique, field
subdwarfs ages, and the helium enrichment ratio $\Delta Y/\Delta Z$
derived from the width of the subdwarfs MS.  These three quantities are all of
paramount importance for cosmological and Galactic evolution issues.

The necessity of this analysis was prompted by the recognition that
isochrones for MS subdwarfs and GC with direct spectroscopical
determinations of the metallicity are not the same if diffusion is
taken properly into account; moreover, differences in the MS location
and TO position between subdwarfs standard and diffusive isochrones
are metallicity and age dependent.

We have considered the {\sl full} effect of atomic diffusion, without
any allowance for possible hydrodynamical mixing phenomena which could
reduce the efficiency of diffusion in Population II stars, as some
observations appear to suggest (see the discussion in Sect. 1).

Our main results are:

\noindent
i) $\alpha$-enhanced calibrated diffusive isochrones reproduce well
within the observational errors the position in the $M_{\rm V}-T_{\rm eff}$
plane of metal poor field subdwarfs with accurate parallaxes and
empirical values of $T_{\rm eff}$ (from the IRFM method), for reasonable
assumptions about their age.
However, with the observational sample considered here
and taking into account
the existing observational uncertainties,
it is impossible from this comparison to demonstrate that diffusion is fully
efficient in Population II stars.

\noindent
ii) MS-fitting distances obtained using current samples of Hipparcos
subdwarfs and standard isochrones are negligibly affected by
atomic diffusion.

\noindent
iii) The estimated subdwarfs ages and age dispersion are strongly
modified when diffusion is properly considered for the subdwarfs
isochrones (absolute ages are significantly reduced). Since the actual
metallicity of subdwarfs in the TO region can be very different 
from the initial one, one must take into account this effect
when deriving age-metallicity relations.

\noindent
iv) The value of $\Delta Y/\Delta Z$ in the Galactic Halo metallicity
range turns out to be systematically underestimated (by
$\delta$($\Delta Y/\Delta Z$)$\approx$1-2) if standard isochrones are
employed.

\section*{Acknowledgments}
We thank H. Ritter for a preliminary reading of the manu\-script
and the referee, Dr. Y. Lebreton, for valuable comments
which improved the presentation of our results.

{}

\end{document}